\documentclass[conference]{IEEEtran}
\usepackage[pscoord]{eso-pic}\newcommand{\placetextbox}[3]{\setbox0=\hbox{#3}\AddToShipoutPictureFG*{\put(\LenToUnit{#1\paperwidth},\LenToUnit{#2\paperheight}){\vtop{{\null}\makebox[0pt][c]{#3}}}}}\placetextbox{.23}{0.055}{979-8-3195-4420-9/26/\$31.00 \textcopyright\ 2026 IEEE}
\IEEEoverridecommandlockouts
\usepackage{cite}
\usepackage[font=footnotesize,skip=3pt]{caption}
\usepackage{amsmath,amssymb,amsfonts}
\usepackage{bbm}
\usepackage{algorithmic}
\usepackage{graphicx}
\usepackage{subcaption}
\usepackage{booktabs}
\usepackage{array}
\usepackage{multirow}
\usepackage{textcomp}
\usepackage{xcolor}
\usepackage{comment}
\usepackage{siunitx}
 \usepackage[font=footnotesize]{caption}
\usepackage[colorlinks=true, allcolors=blue]{hyperref}
\usepackage{orcidlink}
\usepackage[compact]{titlesec}
\titleformat{\paragraph}[runin]
  {\normalfont\normalsize\itshape}{}{0pt}{}
\titlespacing*{\section}{0pt}{6pt plus 2pt minus 2pt}{3pt}
\titlespacing*{\subsection}{0pt}{5pt plus 2pt minus 1pt}{2pt}
\titlespacing*{\subsubsection}{0pt}{4pt plus 1pt minus 1pt}{2pt}
\titlespacing*{\paragraph}{0pt}{3pt plus 1pt minus 1pt}{0.5em}

\usepackage{enumitem}
\setlist{noitemsep, topsep=2pt, parsep=2pt, partopsep=0pt}

\def\BibTeX{{\rm B\kern-.05em{\sc i\kern-.025em b}\kern-.08em
    T\kern-.1667em\lower.7ex\hbox{E}\kern-.125emX}}

\begin{document}

\bstctlcite{BSTcontrol}

\title{Retrieval-Based Cross-Domain Generalization in Optical Networks via Global Features\thanks{This work has been supported by the EUREKA CELTIC-NEXT SUSTAINET-Advance project, funded by the Swiss Innovation Agency Innosuisse No. 119.588 INT-ICT, and by Vinnova (Sweden's Innovation Agency) No. 2025-02987. Corresponding author: ali.alhousseini@supsi.ch }}
\author{
\IEEEauthorblockN{
Ali Al Housseini~\orcidlink{0009-0003-6682-474X}\IEEEauthorrefmark{1}\IEEEauthorrefmark{2},
Carlos Natalino~\orcidlink{0000-0001-7501-5547}\IEEEauthorrefmark{3},
Tiziano Leidi~\orcidlink{0000-0002-6335-7977}\IEEEauthorrefmark{1}
Paolo Monti~\orcidlink{0000-0002-5636-9910}\IEEEauthorrefmark{3},
Omran Ayoub~\orcidlink{0000-0002-3884-3594}\IEEEauthorrefmark{1}
}
\IEEEauthorblockA{\IEEEauthorrefmark{1}
University of Applied Sciences and Arts of Southern Switzerland, Lugano, Switzerland}
\IEEEauthorblockA{\IEEEauthorrefmark{2}
University of Southern Switzerland, Lugano, Switzerland}
\IEEEauthorblockA{\IEEEauthorrefmark{3}
Department of Electrical Engineering, Chalmers University of Technology, Gothenburg, Sweden}

}

\maketitle

\begin{abstract}
We propose a retrieval-based framework for cross-domain quality-of-transmission (QoT) estimation that leverages transferable feature representations while avoiding reliance on source-domain-specific decision boundaries. The proposed approach supports both zero-shot and few-shot adaptation without requiring model retraining. Experimental results on cross-domain QoT datasets demonstrate improved generalization performance compared with conventional machine learning baselines and recent contrastive learning approaches, highlighting the potential of retrieval-based inference for robust optical network automation.
\end{abstract}

\section{Introduction}
Machine learning (ML) techniques are increasingly being explored to support data-driven decision-making in optical networks, with applications spanning lightpath quality of transmission (QoT) estimation, fault management, and resource allocation \cite{Gao_2024_GeneralizationCognitiveOptical, Sadighi_2025_GeneralizabilityMLBasedClassification, Liu_2025_ModuleEnhanceGeneralization}. However, ML models are typically trained under specific network conditions, operational configurations, and topological settings. In deployment, these conditions may no longer hold due to network evolution, changes in traffic patterns, or differences in topology and operating regimes. As a result, the statistical properties of the input data may diverge from those observed during training, leading to a phenomenon known as distribution shift. Such shifts can significantly degrade the predictive performance and reliability of ML models, limiting their ability to generalize across heterogeneous scenarios \cite{usmani2022transfer, usmani2024integrating, aladin2025automated}.

This cross-domain generalization challenge represents a major barrier to the operational adoption of ML in optical networks. Indeed, maintaining model performance across diverse network environments often requires retraining on representative target-domain data, resulting in costly and time-consuming data collection and labeling efforts. 

Several approaches have been proposed to mitigate the impact of distribution shifts and improve cross-domain generalization \cite{usmani2022transfer, usmani2024integrating, zhou2023neuron, zhou2024evolutionary}. Transfer learning and fine-tuning adapt models to a target domain by leveraging knowledge acquired from a source domain; however, their effectiveness often depends on the availability of representative labeled target-domain data and may deteriorate under limited-data regimes due to overfitting or negative transfer. Similarly, contrastive and metric-learning approaches aim to learn transferable or domain-invariant embeddings, but when these embeddings are coupled to a parametric classifier, the final decision boundary may still encode source-domain-specific structure \cite{alhousseini2026ondm}. Consequently, there remains a need for learning frameworks that can exploit transferable representations while reducing reliance on target-domain labels and minimizing sensitivity to source-domain-specific classification boundaries \cite{alhousseini2026ondm, usmani2022transfer, Natalino_2026_UnifiedSiameseLearning}.

In this work, we propose a retrieval-based classification framework for cross-domain QoT estimation that eliminates the need for a parametric classifier during inference. The proposed approach builds upon the cross-domain representation learning framework of \cite{alhousseini2026ondm}, while replacing the conventional classifier with a nearest-neighbor inference mechanism. The key intuition is that, although contrastive learning can produce transferable feature representations, the decision boundaries learned by a source-trained classifier may still limit generalization under distribution shifts. To address this limitation, our proposed approach performs classification directly in the learned embedding space through similarity-based retrieval. More specifically, a $\mathcal{G}$-network maps each lightpath feature vector to a global embedding space optimized using a multi-similarity objective. After training, source-domain embeddings and their corresponding labels are stored in a reference database. During inference, a target-domain sample is projected into the same embedding space by the frozen $\mathcal{G}$-network, and its nearest reference samples are retrieved using cosine similarity. The final QoT label is then determined by majority voting among the top-$k$ retrieved neighbors. Thus, the prediction is determined by the neighborhood structure of the learned embedding space rather than by a fixed classifier trained only on the source domain. This can improve generalization because the decision is made based on local similarity relationships in the contrastive embedding space, rather than from a global decision boundary fitted to the source distribution.


\begin{figure*}[t]
    \centering
    \includegraphics[width=0.99\linewidth]{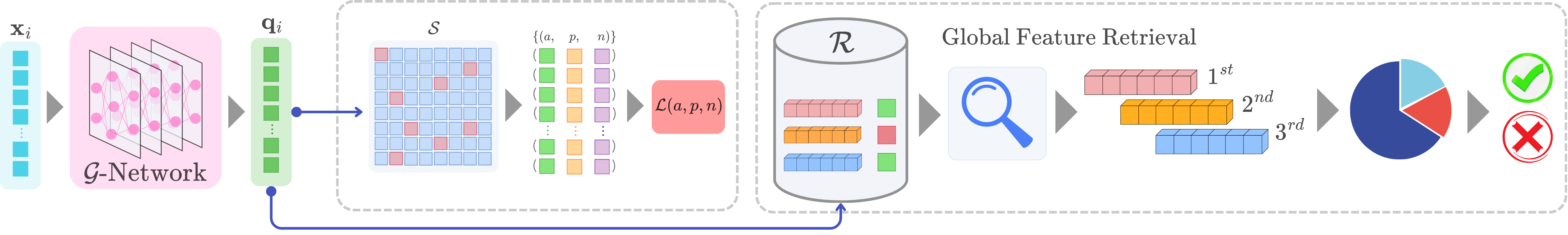}
    \caption{\footnotesize{Proposed retrieval-based contrastive QoT estimation pipeline. 
Each input lightpath feature vector $\mathbf{x}_i$ is mapped by the $\mathcal{G}$-network to a global embedding $\mathbf{q}_i$. 
During training, pairwise similarities within each mini-batch form the similarity matrix $\mathcal{S}$, from which informative triplets are mined to optimize the contrastive loss $\mathcal{L}_{CL}$. 
After training, source-domain embeddings and labels are stored in the reference database $\mathcal{R}$. 
At inference time, a query embedding is compared with the database using cosine similarity, the top-$k$ nearest neighbors are retrieved, and the final QoT label is assigned by majority voting.}}
    \label{fig:fullpipeline}
\end{figure*}

The main contributions of this paper can be summarized as follows:
\begin{itemize}
\item We propose a retrieval-based framework for cross-domain QoT estimation that replaces the conventional source-trained parametric classifier with similarity-based inference in a learned contrastive embedding space, thereby reducing sensitivity to source-domain-specific decision boundaries under distribution shifts.

\item We introduce a lightweight few-shot adaptation mechanism based on database augmentation, whereby a small number of labeled target-domain samples can be incorporated into the retrieval database without retraining or fine-tuning the embedding model.

\item We conduct a comprehensive cross-domain evaluation using QoT datasets generated under distinct optical network configurations and class distributions. The results demonstrate that the proposed retrieval-based inference strategy consistently outperforms strong tabular learning baselines and recent contrastive learning approaches, while further gains can be achieved through the proposed few-shot adaptation mechanism.
\end{itemize}



\section{Related Work}
Cross-domain generalization in optical networks has been addressed from several complementary directions. Empirical studies have quantified the severity of the problem: ML classifiers for State-of-Polarization-based event detection can drop from over 98\% to under 9\% accuracy across spectral bands \cite{Sadighi_2025_GeneralizabilityMLBasedClassification}, and OSNR regression models fail entirely in zero-shot multi-organization settings \cite{Akbari_2025_LeveragingSharedData}.
Transfer learning (TL) has been widely explored as a remedy, including weight-based knowledge transfer and knowledge distillation for GSNR estimation \cite{usmani2022transfer, usmani2024integrating}, neuron-level fine-tuning via importance ranking \cite{zhou2023neuron, zhou2024evolutionary}, margin-driven update triggers \cite{Lechowicz_2025_QoTEstimationMarginDriven}, and feature-selection-based domain adaptation \cite{aladin2025automated}.
Architectural modifications have also been proposed, such as modular component decomposition \cite{Gao_2024_GeneralizationCognitiveOptical}, lightweight adapter modules \cite{Liu_2025_ModuleEnhanceGeneralization}, and cascaded trainable loss models \cite{Wang_2024_MultiSpanOpticalPower}.
However, TL methods require explicit fine-tuning on target-domain data and are susceptible to negative transfer under small-sample conditions, while architectural approaches typically require structural re-mapping for each new network configuration.

Representation learning through metric learning offers a more principled path toward domain-aware features.
Siamese networks have been applied to modulation format recognition \cite{Natalino_2019_OneshotLearningModulation}, fault root-cause analysis \cite{Gao_2024_FaultTracingBased}, and anomaly detection and classification \cite{Natalino_2026_UnifiedSiameseLearning}, demonstrating the value of similarity-based reasoning in optical networks.
In \cite{alhousseini2026ondm}, joint contrastive and classification learning was shown to produce embeddings that generalize across heterogeneous QoT domains; however, the final prediction still relies on a parametric projection head whose decision boundary is learned from the source distribution.

Our work departs from this direction by using the learned contrastive representation directly as a retrieval space.
Instead of attaching a classifier to the embedding, the proposed method classifies each query through nearest-neighbor retrieval and majority voting in the global embedding space.
This changes the role of representation learning: the embedding is not only an intermediate input to a classifier, but the space in which non-parametric inference is performed.
In other words, an optical network operator can reuse the trained embedding model as a similarity engine: new validated lightpaths can be added to the reference database as operational knowledge becomes available, without redesigning or retraining the model for every deployment scenario.
Because the retrieval mechanism only requires embedded samples and labels, the approach is also task-agnostic and can be applied beyond binary QoT estimation to other optical-network classification tasks in which similarity among operational examples is meaningful.
The resulting framework avoids fitting a source-domain parametric decision boundary and enables few-shot adaptation by inserting labeled target samples into the retrieval database without updating the neural network weights.
To our knowledge, this is the first investigation of retrieval-based classification over contrastive QoT embeddings for cross-domain optical-network generalization.

\section{Retrieval-based Contrastive Learning}

\subsection{Problem Definition}
We address cross-domain QoT classification under domain shift. 
Let $\mathcal{D}_{src}=\{(\mathbf{x}_i,y_i)\}_{i=1}^{N_{src}}$ denote a labeled source-domain dataset with $(\mathbf{x}_i,y_i)\sim\mathcal{P}$, and let $\mathcal{D}_{tgt}=\{(\mathbf{x}_j,y_j)\}_{j=1}^{N_{tgt}}$ denote a target-domain dataset with $(\mathbf{x}_j,y_j)\sim\mathcal{Q}$, where $\mathcal{P}\neq\mathcal{Q}$. 
Here, $\mathbf{x}_i\in\mathbb{R}^{d}$ is the lightpath feature vector and $y_i\in\{0,1\}$ is the QoT class label. 
The shift between $\mathcal{P}$ and $\mathcal{Q}$ may arise from differences in topology, transceiver configuration, link-length distribution, physical-layer impairments, traffic loading, and class prior. 
The learning objective is to train on $\mathcal{D}_{src}$ and obtain reliable predictions on $\mathcal{D}_{tgt}$ without assuming that labeled target-domain data are available during training.

Instead of optimizing a classifier that directly maps $\mathbf{x}_i$ to $y_i$, the proposed method learns a representation space in which QoT-relevant similarity can be measured across domains. 
The central hypothesis is that a contrastively learned embedding can preserve task-relevant neighborhood structure more robustly than a source-trained parametric decision boundary. 
Classification is therefore deferred to a retrieval stage, described in Section~\ref{subsec:retrieval}, rather than performed by a learned projection head.

\subsection{Methodology}

Fig.~\ref{fig:fullpipeline} illustrates the proposed retrieval-based contrastive QoT estimation pipeline. 
The architecture contains a single neural feature extractor, denoted as the $\mathcal{G}$-network, which maps each lightpath feature vector $\mathbf{x}_i$ to a global embedding
\begin{equation}
\mathbf{q}_i=f_{\mathcal{G}}(\mathbf{x}_i)\in\mathbb{R}^{d_g}.
\end{equation}
During training, the embedding space is shaped using a contrastive objective so that lightpaths with the same QoT class are encouraged to be close, while lightpaths from different classes are pushed apart. 
After training, the network is frozen, and source-domain embeddings are stored together with their labels in a reference database. 
At inference time, a query lightpath is embedded by the frozen $\mathcal{G}$-network, compared with the database using cosine similarity, and assigned a label by majority voting over its top-$k$ nearest neighbors.

This design separates representation learning from the final decision rule. 
The neural network learns a metric space, while the final classification is non-parametric and depends on the local neighborhood of the query in that space. 
This is a key difference from the joint contrastive-classification framework in~\cite{alhousseini2026ondm}, where the embedding is followed by a parametric projection head trained with cross-entropy. 
In the proposed method, no such classifier head is used.

\subsection{Contrastive Embedding Learning}

The $\mathcal{G}$-network is a deep neural backbone 
$f_{\mathcal{G}}:\mathbb{R}^{d}\rightarrow\mathbb{R}^{d_g}$ trained on the source dataset $\mathcal{D}_{src}$. 
For a mini-batch of size $B$, the network produces embeddings $\{\mathbf{q}_i\}_{i=1}^{B}$. 
Pairwise cosine similarities are computed to form the similarity matrix $\mathcal{S}$:
\begin{equation}
\mathcal{S}_{ij}=
\frac{\mathbf{q}_i^\top\mathbf{q}_j}
{\|\mathbf{q}_i\|_2\|\mathbf{q}_j\|_2}.
\label{eq:similarity_matrix}
\end{equation}

The model is trained using the class-conditional sampler and Multi-Similarity miner adopted in~\cite{alhousseini2026ondm}. 
The sampler mitigates class imbalance by ensuring that each mini-batch contains samples from both QoT classes, while the miner selects informative positive and negative relations rather than treating all within-batch pairs equally. 
The Multi-Similarity loss is
\begin{equation}
\begin{aligned}
\mathcal{L}_{CL}
&= \frac{1}{B} \sum_{i=1}^{B} \Bigg\{
\frac{1}{\alpha} \log\Bigg[1 + \sum_{j\in \mathcal{P}_i} 
e^{-\alpha(\mathcal{S}_{ij}-m)}\Bigg] \\
&\qquad\qquad
+ \frac{1}{\beta} \log\Bigg[1 + \sum_{k\in \mathcal{N}_i} 
e^{\beta(\mathcal{S}_{ik}-m)}\Bigg]
\Bigg\},
\label{eq_loss}
\end{aligned}
\end{equation}
where $\mathcal{P}_i$ and $\mathcal{N}_i$ denote the mined positive and negative sets associated with anchor $i$, respectively. 
The parameters $\alpha$, $\beta$, and $m$ control the pair-weighting strength and similarity margin~\cite{wang2019multi}.

The role of $\mathcal{L}_{CL}$ is instead to construct a discriminative metric space in which nearest-neighbor retrieval becomes meaningful for QoT classification.

\subsection{Retrieval-Based Classification}
\label{subsec:retrieval}

After training, the $\mathcal{G}$-network is frozen, and the source-domain samples are encoded into a reference database
\begin{equation}
\mathcal{R}=
\{(\mathbf{q}_r,y_r):\mathbf{q}_r=f_{\mathcal{G}}(\mathbf{x}_r),
(\mathbf{x}_r,y_r)\in\mathcal{D}_{src}\}.
\label{eq:reference_database}
\end{equation}
For a lightpath input $\mathbf{x}_i$, the frozen network produces $\mathbf{q}_i=f_{\mathcal{G}}(\mathbf{x}_i)$. 
The query is compared with each database embedding using cosine similarity:
\begin{equation}
s(i,r)=
\frac{\mathbf{q}_i^\top\mathbf{q}_r}
{\|\mathbf{q}_i\|_2\|\mathbf{q}_r\|_2}.
\label{eq:cosine_retrieval}
\end{equation}
The $k$ database samples with the largest similarity values define the retrieval neighborhood
\begin{equation}
\mathcal{H}_k(i)=
\operatorname{arg\,topk}_{r:(\mathbf{q}_r,y_r)\in\mathcal{R}} s(i,r).
\label{eq:topk}
\end{equation}
The predicted QoT label is obtained by majority voting:
\begin{equation}
\hat{y}_i=
\operatorname{mode}\{y_r:r\in\mathcal{H}_k(i)\}.
\label{eq:retrieval_vote}
\end{equation}
For binary classification and odd $k$, no tie-breaking rule is required.

This is not equivalent to applying $k$-nearest neighbors directly on raw tabular lightpath features. 
Raw features may contain heterogeneous scales, topology-specific correlations, and variable importance that changes across domains. 
Here, nearest-neighbor inference is performed in a contrastively learned space where the geometry is explicitly optimized to reflect QoT class relationships. 
The final decision, therefore, depends on the learned neighborhood structure of lightpaths rather than on a source-trained classifier boundary.

\subsection{Training and Database Adaptation}

The full procedure consists of three steps. 
First, $f_{\mathcal{G}}$ is trained on $\mathcal{D}_{src}$ using Eq.~\eqref{eq_loss}. 
Second, the trained network is frozen and the reference database $\mathcal{R}$ in Eq.~\eqref{eq:reference_database} is populated with source-domain embeddings and labels. 
Third, inference is performed by cosine-similarity retrieval and majority voting according to Eqs.~\eqref{eq:cosine_retrieval}--\eqref{eq:retrieval_vote}.

When a small labeled target-domain subset $\mathcal{A}\subset\mathcal{D}_{tgt}$ becomes available, adaptation is performed by database augmentation:
\begin{equation}
\mathcal{R}^{+}
=
\mathcal{R}
\cup
\{(f_{\mathcal{G}}(\mathbf{x}_a),y_a):(\mathbf{x}_a,y_a)\in\mathcal{A}\}.
\label{eq:database_augmentation}
\end{equation}
No network weights are updated. 
Thus, incorporating target-domain knowledge requires only embedding extraction and database insertion, whereas parametric alternatives require fine-tuning or retraining under scarce target supervision. 
This distinction is operationally important for early deployment, where a small number of validated target lightpaths may be available but insufficient to reliably train or calibrate a new neural classifier.

\section{Experimental Results}

\subsection{Experimental Setup}
We evaluate cross-domain generalization using two public datasets from the Fraunhofer HHI QoT dataset collection~\cite{fraunhofer_qot_collection}, fully described in~\cite{bergk2022qotdataset}. We denote Dataset~01 and Dataset~02 as $D1$ and $D2$, respectively. 
Both datasets are generated over the CONUS topology and include 100/200/400~Gb/s lightpath requests, but they differ in transceiver configuration and has 24.2\% infeasible lightpaths, whereas $D2$ uses Predefined Transceiver Mode and has only 7.3\% infeasible lightpaths.

We consider both transfer directions, $D1\rightarrow D2$ and $D2\rightarrow D1$. 
Two evaluation regimes are studied, namely, \emph{zero-shot} and \emph{few-shot}.  
In the zero-shot regime, the model is trained only on the source dataset and evaluated directly on the target dataset, without using any labeled target samples for training or database construction. 
For the proposed method, this means that the reference database $\mathcal{R}$ contains only source-domain embeddings. 
In the few-shot regime, a small fraction of labeled target samples is made available after source-domain training. 
For the proposed retrieval classifier, these samples are encoded by the frozen $\mathcal{G}$-network and inserted into $\mathcal{R}$; no network weights are updated. 

All preprocessing parameters are fitted on the source training set and then applied unchanged to the target domain. 
The proposed method classifies target samples by cosine-similarity retrieval in the learned contrastive embedding space, followed by majority voting over the top-$k=3$ retrieved neighbors. 
The few-shot analysis reports performance as the percentage of labeled target samples increases from 1\% to 4\%, allowing us to assess whether adaptation is stable in the low-label regime.

We compare the proposed contrastive retrieval classifier against competitive tabular-learning baselines, namely Random Forest (RF), ExtraTrees (Extr), Logistic Regression (LR), XGBoost (XGB), and CatBoost (CatB). 
We also include the joint contrastive-classification learning method (JCCL) from~\cite{alhousseini2026ondm}, which uses the same representation-learning backbone but performs the final decision with a learned parametric projection head. 
This comparison isolates the effect of replacing the source-trained classifier with nearest-neighbor retrieval over the contrastive embedding space.

Performance is reported using Accuracy (Acc), Macro-F1 (MF1), PR-AUC, and ROC-AUC. Acc is included for comparability with prior QoT-estimation studies, but it is not sufficient on its own because the datasets are class-imbalanced. MF1 is used as the primary hard-decision metric, since it assigns equal weight to the healthy and failure classes. PR-AUC and ROC-AUC are reported as threshold-independent ranking metrics, with the failure class treated as the event of interest; PR-AUC is particularly relevant because failures correspond to the minority and operationally critical class.

For parametric baselines, the ranking score is the predicted probability or decision score assigned to the failure class. 
For the proposed retrieval classifier, hard labels are obtained by majority voting over the top-$k$ retrieved neighbors. 
The corresponding ranking score is defined as the fraction of retrieved neighbors labeled as unacceptable lightpaths:
\begin{equation}
\rho_i=
\frac{1}{k}
\sum_{r\in\mathcal{H}_k(i)}
\mathbf{1}\!\left[y_r=c_{\mathrm{fail}}\right],
\label{eq:knn_score}
\end{equation}
where $c_{\mathrm{fail}}$ denotes the failure class. 
This score is the local empirical failure density in the learned embedding space and is used to compute PR-AUC and ROC-AUC for the retrieval classifier.

\begin{table}[t]
\centering
\caption{\footnotesize Cross-dataset evaluation for D1$\leftrightarrow$D2 transfer. Best result per column in \textbf{bold}; second-best \underline{underlined}.}
\label{tab:transfer_d1_d2_only}
\scriptsize
\setlength{\tabcolsep}{3pt}
\begin{tabular}{l | cccc | cccc}
\toprule
 & \multicolumn{4}{c|}{\textbf{D1 $\rightarrow$ D2}} 
 & \multicolumn{4}{c}{\textbf{D2 $\rightarrow$ D1}} \\
\cmidrule{2-5} \cmidrule{6-9}
Model 
& Acc & MF1 & PR-AUC & ROC-AUC 
& Acc & MF1 & PR-AUC & ROC-AUC \\
\midrule
RF   
& 0.593 & 0.515 & 0.552 & 0.594
& 0.981 & 0.980 & 0.971 & 0.973 \\

Extr 
& 0.575 & 0.487 & 0.542 & 0.578
& 0.962 & 0.940 & 0.956 & 0.971 \\

LR   
& 0.550 & 0.495 & 0.532 & 0.550
& 0.620 & 0.612 & 0.818 & 0.680 \\

XGB  
& 0.602 & 0.534 & 0.542 & 0.578
& \underline{0.986} & 0.982 & 0.972 & \underline{0.985} \\

CatB 
& 0.604 & 0.532 & 0.558 & \underline{0.604}
& 0.985 & \underline{0.985} & \underline{0.984} & 0.975 \\


JCCL \cite{alhousseini2026ondm} 
& \underline{0.674} & \underline{0.644} & \underline{0.656} & 0.592
& 0.982 & 0.979 & 0.973 & 0.977 \\

\midrule
Ours 
& \textbf{0.725} & \textbf{0.743} & \textbf{0.681} & \textbf{0.698}
& \textbf{0.993} & \textbf{0.988} & \textbf{0.988} & \textbf{0.991} \\

\bottomrule
\end{tabular}
\end{table}

\begin{figure}[t]
    \centering
    \includegraphics[width=1\linewidth]{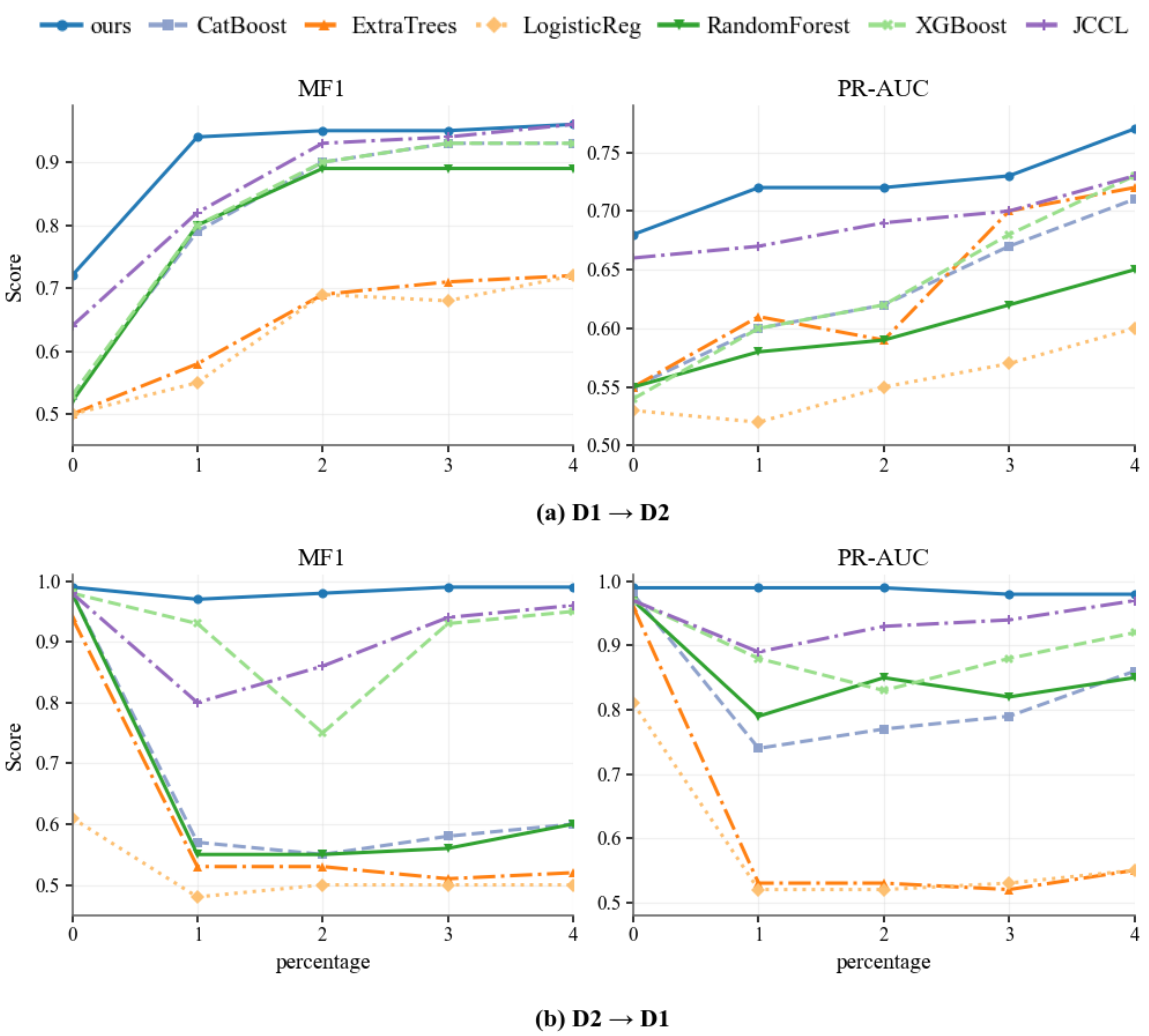}
    \caption{Few-shot adaptation results for (a) $D1\rightarrow D2$ and (b) $D2\rightarrow D1$. 
Each curve shows MF1 or PR-AUC as a function of the percentage of labeled target-domain samples made available after source-domain training.}
\label{fig:fewshot_adaptation}
\end{figure}

\subsection{Numerical Results and Discussion}

\textbf{Zero-Shot Cross-Domain Results.} Table~\ref{tab:transfer_d1_d2_only} reports the performance of the various approaches under the zero-shot regime for $D1\rightarrow D2$ and $D2\rightarrow D1$.

The zero-shot results show that the proposed retrieval-based classifier provides the strongest MF1 and PR-AUC in both transfer directions. 
For $D1\rightarrow D2$, it reaches $0.743$ MF1 and $0.681$ PR-AUC, improving over the best classical baselines by $0.209$ and $0.123$, respectively, and over JCCL by $0.099$ and $0.025$. 
For $D2\rightarrow D1$, it obtains $0.988$ MF1 and $0.988$ PR-AUC, improving over the best classical baselines by $0.003$ and $0.004$, and over JCCL by $0.009$ and $0.015$. 
These gains show that the improvement is not only due to contrastive representation learning, but also to using the learned embedding directly as a retrieval space rather than as input to a source-trained classifier. 
\paragraph*{Takeaway-1:} Neighborhood-based inference over the contrastive embedding improves the key metrics in both transfer directions. 

The larger gain occurs for $D1\rightarrow D2$, where the source and target domains differ substantially in transceiver mode and class imbalance. 
In this setting, the classical baselines achieve only $0.534$ MF1 and $0.558$ PR-AUC, while JCCL achieves $0.644$ MF1 and $0.656$ PR-AUC. 
The reverse direction, $D2\rightarrow D1$, is less challenging for tree-based baselines, with CatBoost already achieving $0.985$ MF1 and $0.984$ PR-AUC.  Nevertheless, the proposed retrieval classifier still obtains the best values, with $0.988$ MF1 and $0.988$ PR-AUC. 
\paragraph*{Takeaway-2:} Retrieval brings the largest benefit when source-trained decision boundaries are poorly aligned with the target distribution, while preserving performance in the easier direction.

\textbf{Few-Shot Adaptation Results.} Fig.~\ref{fig:fewshot_adaptation} reports the performance of the various approaches under the few-shot in terms of MF1 adn PR-AUC for both transfer directions. 
For the proposed method, adaptation is performed by inserting the labeled target embeddings into the reference database while keeping the $\mathcal{G}$-network frozen. 
Thus, the curves measure the effect of enriching the retrieval database rather than retraining the model.

For $D1\rightarrow D2$, the proposed method benefits immediately from target-domain support. 
MF1 increases sharply with only $1\%$ labeled target samples and remains high as more samples are added. 
This indicates that a small number of target embeddings is sufficient to reshape the local neighborhoods used by the retrieval classifier. 
The PR-AUC curve also improves steadily, showing that the added target samples not only improve hard-label decisions, but also improve the ranking of failure samples in the retrieval space. JCCL also improves with target data and approaches the proposed method at larger injection ratios, but the retrieval approach is stronger in the low-data regime, where adaptation stability is most important.


For $D2\rightarrow D1$, the proposed method starts from a high-performance operating point and remains stable across all injection ratios.
MF1 stays close to saturation, and PR-AUC remains near the top throughout the adaptation process. 
This stability is important: adding a small number of labeled target samples does not degrade the retrieval classifier. 
In contrast, several baselines exhibit pronounced non-monotonic behavior after the first injection step, especially in MF1 and PR-AUC. 
This suggests that small target subsets may be insufficiently representative for reliable retraining, causing temporary degradation before recovery at higher injection ratios.

In conclusion, Fig.~\ref{fig:fewshot_adaptation} supports the motivation of the proposed framework. 
In early deployment, operators may be able to validate a small number of target-domain lightpaths, but not enough to reliably retrain a classifier. 
The proposed method can exploit such limited supervision through database augmentation, improving or preserving MF1 and PR-AUC without updating network weights. This leads to the last takeaway.

\paragraph*{Takeaway-3:} This behavior is important for real-world deployment: the proposed method adapts by accumulating validated target lightpaths in the retrieval database, which improves coverage of the target operating regime without perturbing the learned embedding. This makes adaptation more robust and operationally simpler than retraining parametric models whenever a small amount of new target-domain data becomes available.



\section{Conclusion}
We propose a retrieval-based framework for cross-domain generalization of lightpath QoT estimation in optical networks. Experiments on cross-domain transfer show that the proposed approach improves generalization over classical tabular baselines and previous state-of-the-art methods. The approach also enables simple few-shot adaptation by inserting labeled target samples into the retrieval database without updating network weights. 

\bibliographystyle{IEEEtran}
\bibliography{references}

@IEEEtranBSTCTL{BSTcontrol,
  CTLuse_forced_etal       = "yes",
  CTLmax_names_forced_etal = "1",
  CTLnames_show_etal       = "1"
}

@inproceedings{alhousseini2026ondm,
  author  = {Ali Al Housseini and Carlos Natalino and Paolo Monti and Omran Ayoub},
  title = {{C}ross-{D}omain {G}eneralization in {O}ptical {N}etworks via {J}oint {C}ontrastive and {C}lassification {L}earning},
  booktitle = {International Conference on Optical Network Design and Modeling (ONDM)},
  year = {2026},
}

@article{Gao_2024_GeneralizationCognitiveOptical,
  author  = {Hanyu Gao and Xiaoliang Chen and Chao Lu and Zhaohui Li},
  title   = {On the generalization of cognitive optical networking applications using composable machine learning},
  journal = {Journal of Optical Communications and Networking},
  year    = {2024},
}

@inproceedings{Natalino_2019_OneshotLearningModulation,
  author       = {Carlos Natalino and Aleksejs Udalcovs and Lena Wosinska and Oskars Ozolins and Marija Furdek},
  title        = {One-Shot Learning for Modulation Format Identification in Evolving Optical Networks},
  booktitle    = {OSA Advanced Photonics Congress (AP)},
  year         = {2019},
}

@article{Liu_2025_ModuleEnhanceGeneralization,
  author  = {Zheng Liu and Tiegen Liu and Jian Zhao and Joshua Uduagbomen and Yulin Wang and Sergei Popov and Tianhua Xu},
  title   = {A Module to Enhance the Generalization Ability of End-to-End Deep Learning Systems in Optical Fiber Communications},
  journal = {Journal of Lightwave Technology},
  year    = {2025},
}

@article{aladin2025automated,
  author    = {Sandra Aladin and Lena Wosinska and Christine Tremblay},
  title     = {Automated, Interpretable and Efficient {ML} Models for Real-World Lightpaths' Quality of Transmission Estimation},
  journal   = {IEEE Open Journal of the Communications Society},
  year      = {2025},
}

@article{zhou2024evolutionary,
  author  = {Yuhang Zhou and Zhiqun Gu and Jiawei Zhang and Yuefeng Ji},
  title   = {Evolutionary neuron-level transfer learning for {QoT} estimation in optical networks},
  journal = {Journal of Optical Communications and Networking},
  year    = {2024},
}

@article{usmani2024integrating,
  author  = {Fehmida Usmani and Ihtesham Khan and Arsalan Ahmad and Vittorio Curri},
  title   = {Integrating Knowledge Distillation and Transfer Learning for Enhanced {QoT}-Estimation in Optical Networks},
  journal = {IEEE Access},
  year    = {2024},
}

@inproceedings{usmani2022transfer,
  author       = {Fehmida Usmani and Ihtesham Khan and Muhammad Umar Masood and Arsalan Ahmad and Muhammad Shahzad and Vittorio Curri},
  title        = {Transfer Learning Aided {QoT} Computation in Network Operating with the {400ZR} Standard},
  booktitle    = {International Conference on Optical Network Design and Modeling (ONDM)},
  year         = {2022},
}

@inproceedings{zhou2023neuron,
  author    = {Yuhang Zhou and Zhiqun Gu and Jiawei Zhang and Yuefeng Ji},
  title     = {Neuron-level Transfer Learning for {ANN}-based {QoT} Estimation in Optical Networks},
  booktitle = {Asia Communications and Photonics Conference / International Photonics and Optoelectronics Meetings (ACP/POEM)},
  year      = {2023},
}

@inproceedings{Wang_2024_MultiSpanOpticalPower,
  author       = {Zehao Wang and Yue-Kai Huang and Shaobo Han and Ting Wang and Dan Kilper and Tingjun Chen},
  title        = {Multi-Span Optical Power Spectrum Prediction Using {ML}-based {EDFA} Models and Cascaded Learning},
  booktitle    = {Optical Fiber Communication Conference (OFC)},
  year         = {2024},
}

@inproceedings{Sadighi_2025_GeneralizabilityMLBasedClassification,
  author    = {Leyla Sadighi and Carlos Natalino and Stefan Karlsson and Marco Ruffini and Eoin Kenny and Lena Wosinska and Marija Furdek},
  title     = {Generalizability of {ML}-Based Classification of State of Polarization Signatures Across Different Bands and Links},
  booktitle = {European Conference on Optical Communications (ECOC)},
  year      = {2025},
}

@inproceedings{Akbari_2025_LeveragingSharedData,
  author    = {Hassan Akbari and Xiao Ma and Behnam Shariati and Pooyan Safari and Angela Mitrovska and Johannes K. Fischer and Stephan Pachnicke and Jasper M{\"u}ller and Ronald Freund},
  title     = {Leveraging Shared Data and Models for {ML}-Based {QoT} Estimation: Toward Standardized and Generalizable Models},
  booktitle = {European Conference on Optical Communications (ECOC)},
  year      = {2025},
}

@inproceedings{Natalino_2026_UnifiedSiameseLearning,
  author    = {Carlos Natalino and Fl{\'a}via Pessoa Monteiro and Paolo Monti},
  title     = {A Unified {Siamese} Learning Framework for Zero-Day Anomaly Detection and Classification in Optical Networks},
  booktitle = {Optical Fiber Communication Conference (OFC)},
  year      = {2026},

}

@inproceedings{Gao_2024_FaultTracingBased,
  author    = {Yuxuan Gao and Bingli Guo and Yu Zhou and Yuting Ma and Kuan Yan and Shanguo Huang},
  title     = {Fault Tracing Based on {Siamese} Neural Network for Optical Networks},
  booktitle = {IEEE Opto-Electronics and Communications Conference (OECC)},
  year      = {2024},
}

@inproceedings{Lechowicz_2025_QoTEstimationMarginDriven,
  author       = {Piotr Lechowicz and Carlos Natalino and Paolo Monti},
  title        = {{QoT} Estimation with Margin-Driven Transfer Learning in Time-Varying Optical Networks},
  booktitle    = {Optical Fiber Communication Conference (OFC)},
  year         = {2025},
}

@inproceedings{wang2019multi,
  author    = {Xun Wang and Xintong Han and Weilin Huang and Dengke Dong and Matthew R. Scott},
  title     = {Multi-Similarity Loss with General Pair Weighting for Deep Metric Learning},
  booktitle = {Proceedings of the IEEE/CVF Conference on Computer Vision and Pattern Recognition},
  year      = {2019},
}

@article{bergk2022qotdataset,
  author  = {Geronimo Bergk and Behnam Shariati and Pooyan Safari and Johannes K. Fischer},
  title   = {{ML}-assisted {QoT} estimation: a dataset collection and data visualization for dataset quality evaluation},
  journal = {Journal of Optical Communications and Networking},
  year    = {2022},
}

@misc{fraunhofer_qot_collection,
  author       = {{Fraunhofer Heinrich-Hertz-Institut}},
  title        = {{QoT} Dataset Collection},
  year         = {2026},
  howpublished = {\url{https://www.hhi.fraunhofer.de/en/pn-software/qot-dataset-collection.html}},
}

\end{document}